\documentclass[]{emulateapj}
\usepackage{graphicx}
\usepackage{hhline}
\usepackage{xcolor}
\usepackage[normalem]{ulem}
\usepackage{amssymb, amsmath}

\newcommand{\nsat}{n_{\rm sat}}
\newcommand{\Ms}{M_{\odot}}

\newcommand{\leff}{\widetilde\Lambda}
\newcommand{\mc}{\mathcal{M}_c}

\begin{document}

\title{Optimized statistical approach for comparing multi-messenger neutron star data}
\author{Carolyn A. Raithel, Feryal \"Ozel, \& Dimitrios Psaltis}
\affiliation{Department of Astronomy and Steward Observatory, University of Arizona, 933 N. Cherry Avenue, Tucson, Arizona 85721, USA}

\begin{abstract}
The neutron star equation of state is now being constrained from a diverse set of multi-messenger data, including gravitational waves from binary neutron star mergers, X-ray observations of the neutron star radius, and many types of laboratory nuclear experiments. These measurements are often mapped to a common domain for comparison with one another or are used to constrain the predictions of theoretical equations of state. We explore here the statistical biases that can arise when such multi-messenger data are compared or combined across different domains. We find that placing Bayesian priors individually in each domain of measurement can lead to biased constraints. We present a new prescription for defining Bayesian priors consistently across different experiments, which will allow for robust cross-domain comparisons. Using the first two binary neutron star mergers as an example, we show that a uniform prior in the tidal deformability can produce inflated evidence for large radii, while a uniform prior in the radius points towards smaller radii. Finally, using this new prescription, we provide a status update on multi-messenger constraints on the neutron star radius.
\end{abstract}

\maketitle

\section{Introduction}
We are now in an era of true multi-messenger constraints on the neutron star equation of state (EOS), with a wealth of new results coming in from electromagnetic observations of astrophysical sources, gravitational wave detections of binary systems, and laboratory-based nuclear experiments.

On the astrophysical side, X-ray observations of surface emission from neutron stars in low-mass X-ray binaries (LMXBs) have constrained the radii of at least a dozen sources (\citealt{Ozel2009a,Guver2010,Guillot2013,Guillot2014,Heinke2014,Nattila2016,
Ozel2016a,Bogdanov2016}; for a recent review, see \citealt{Ozel2016}). Under the assumption that all neutron stars have a common radius, these measurements combine to yield a narrowly-constrained radius of $R=10.3\pm0.5$~km \citep{Ozel2016a}. Other analyses of the same set of sources under different theoretical priors lead to a range of $10.6-12.6$~km, at 68\% confidence, for a 1.4~$\Ms$ neutron star \citep{Steiner2013}.  Additionally, the NICER collaboration recently reported the first radius constraint for an isolated X-ray pulsar \citep{Bogdanov2019},f which is quite broad but seems to favor relatively large radii, $R=12.71\substack{+1.14\\-1.19}$~km, for a multi-component, phenomenological set of pulse-profile models \citep{Riley2019,Miller2019}. The LIGO-Virgo collaboration has also now detected two likely binary neutron star mergers. The first event, GW170817, provided strong constraints on the effective tidal deformability of the binary neutron star system, $\leff = 300\substack{+430\\-220}$ \citep{Abbott2017a,Abbott2019}. While there was no strong detection of tidal effects in the second event, GW190425, the masses from this event render it likely to be a second binary neutron star system, which some studies have already used in placing new, multi-messenger constraints on the neutron star EOS \citep{LIGO2020,Dietrich2020,Landry2020}.\footnote{\citet{LIGO2020} does point out that, due to the weak measurement of tidal effects, it remains possible that GW190425 contains at least one black hole. Throughout this paper, we will assume that GW190425 was, in fact, a binary neutron star merger, as is assumed in the majority of the discovery paper \citep{LIGO2020}.} 

In addition to these astrophysical measurements, a wide variety of nuclear experiments have placed complementary constraints on the low-density portion of the EOS. For example, the two-body potential can be constrained from nucleon-nucleon scattering data at energies below 350~MeV and from the properties of light nuclei, which directly informs the EOS at densities near the nuclear saturation density, $\nsat$ \citep{Akmal1998,Morales2002}. Experimental constraints are also often expressed in terms of the nuclear symmetry energy, which characterizes the difference in energy between pure neutron matter and symmetric nuclear matter. The value of the nuclear symmetry energy at $\nsat$ and its slope, $L_0$, have been constrained by fits to nuclear masses; by measurements of the neutron skin thickness, the giant dipole resonance, and electric dipole polarizability of $^{208}$Pb; and by observations of isospin diffusion or multifragmentation in heavy ion collisions (e.g., \citealt{Danielewicz2003,Centelles2009,Roca-Maza2013,Tamii2011,Tsang2012}; see \citealt{Oertel2017} for a recent review).

With this diversity of data, the question then arises of how one might robustly compare the results across the various domains. In this paper, we will provide one self-consistent method for comparing posteriors on the neutron star EOS from different types of experimental data, for any domain in which the prior is defined. This method allows one to directly compare previously-published posteriors on observable neutron star properties. As such, this approach is different from, but complementary to full Bayesian inferences, which attempt to combine multi-messenger data to constrain the parameters of the EOS itself (for further discussion, see~$\S$~\ref{sec:motivation}). For the properties considered in this paper, we will focus specifically on recent constraints from  X-ray observations of the neutron star radius, gravitational waves constraints on $\leff$, and nuclear experiments constraining $L_0$.  As we will show, defining priors self-consistently across the multiple domains is critical for ensuring unbiased constraints, especially when the data are sparse or weakly constraining.

We start in $\S$~\ref{sec:motivation} with an overview of currently-used approaches for cross-domain comparisons of neutron star data and we highlight some common pitfalls of these methods, which motivate this paper. In $\S$\ref{sec:Bayes}, we provide a brief review of Bayesian statistics, in order to define the issues that arise when defining priors across different domains. We then introduce consistent sets of Bayesian priors for the various domains of comparison that are relevant for EOS constraints and we comment on the use of a Jeffreys' prior to solve this problem. In $\S$\ref{sec:transform}, we derive a set of analytic transformation equations that facilitate the mapping of posteriors between any two domains, making use of previously-published mappings between the nuclear symmetry energy and the radius, as well as between the radius and the binary tidal deformability. These transformation functions allow for diverse sets of archival posteriors to be compared self-consistently. In $\S$\ref{sec:GWtransform}, we apply the newly-derived priors to the concrete example of the measured tidal deformability from GW170817 and GW190425. The choice of priors strongly dominates for the weakly-informative GW190425, but for both events, the choice of a uniform prior in the tidal deformability inflates the evidence for large radii, while a uniform prior in $R$ points towards smaller radii. Finally, in $\S$\ref{sec:constraints}, we combine a set of archival posteriors from X-ray observations, the two gravitational wave events, and a recent study using heavy-ion collisions and we present summary constraints on the neutron star radius.

\section{Motivation and past work}
\label{sec:motivation}
Many studies have attempted to address the question of how to combine multi-messenger data to constrain the neutron star EOS. Currently, there are several different approaches that are commonly used in the literature. In this section, we will give an overview of these methods and will highlight that, in the absence of a community consensus on the appropriate prior distributions to assume, the particular choice of priors can significantly affect the resulting constraints. This issue becomes particularly pressing when the data under consideration are sparse or weakly constraining.

One way of combining multi-messenger data is to perform a Bayesian inference, in which a single set of priors is defined in the EOS domain and likelihoods are sampled in all domains of measurements \citep[e.g.,][]{Steiner2010,Wade2014,Steiner2016,Ozel2016a,Raithel2016,Riley2018,Landry2019}. Even though a Bayesian inference is, in principle, robust, there are still choices to be made when setting up the framework. For example, while certain priors are widely agreed upon, such as the requirement for the EOS to maintain thermodynamic stability and for the sound speed to remain subluminal, others, such as which parametrization to use or what priors to place on the parameters of a particular model, are still being debated. Several studies have explored the role of the EOS priors in inferences from X-ray radii \citep[e.g.,][]{Steiner2016} or from gravitational waves \citep[e.g.,][]{Carney2018}, and have found that the result is indeed sensitive to the choice of parametrization. In order to avoid this sensitivity, \citet{Landry2019} recently introduced a non-parametric inference scheme, which uses Gaussian processes trained on a sample of theoretical EOS models. While this approach allows for more direct characterization of the errors of the inferred EOS function, it too requires a choice when defining the priors. In \citet{Landry2019}, the priors are conditioned on a set of published EOS, which may reflect historical trends more than the true range of possible physics.

Additionally, Bayesian inference schemes tend to be computationally expensive. Calculating each likelihood requires an integration of the TOV equations to compute masses, radii, and tidal deformabilities for every EOS sampled within the inference scheme. As a result, it remains common to instead make use of archival (i.e., published or publicly available) posteriors on intermediate parameters, such as $R$ or $\leff$ (see \citealt{Riley2018} for further discussion).  

Archival posteriors can provide a useful consistency check between astrophysical observations and a proposed EOS, at low computational cost. This provides an alternative method for using multi-messenger data to inform EOS theory. For example, many recent EOS analyses  \citep[e.g.,][]{Krastev2019,Lim2019,Blaschke2020,Christian2020,Fattoyev2020,Khanmohamadi2020,Marczenko2020} have compared the predictions of their models to some combination of posteriors on $\Lambda_{1.4}$ from GW170817, posteriors on $R$ from GW170817, as inferred either through a Bayesian inference \citep[e.g., using a spectral EOS, as in][]{Abbott2018} or using quasi-universal relations  \citep[e.g.,][]{Annala2018,De2018,Most2018,Raithel2018,Coughlin2019,Radice2019,Raithel2019a}, and posteriors on $R$ from the recent NICER measurement \citep{Miller2019,Riley2019}.
Other studies have sought to incorporate more directly different types of measurements, combining, for example, nuclear constraints on $L_0$ with the LIGO posteriors on $\leff$  \citep{Malik2018,Carson2019} and additionally with NICER posteriors on $R$ \citep{Zimmerman2020},  or directly comparing constraints from different sources (e.g., \citealt{Raithel2019a} for a comparison of LMXB posteriors and radii inferred from posteriors on $\leff$; or \citealt{Raithel2019b} for a comparison of posteriors on $\leff$ and nuclear constraints on $L_0$).

However, each of the above studies combines sets of posteriors that assume different -- and, as we will show, incompatible -- sets of priors. These mixed priors imply inconsistent assumptions about the universe that cannot be held simultaneously. For example, the radius measurements that come from NICER or LMXBs tend to assume a flat prior in $R$. In contrast, the LIGO posteriors on the tidal deformability of GW170817 assume a flat prior in $\leff$ \citep{Abbott2017a,Abbott2019}, and any radius constraints from GW170817 that were inferred using the quasi-universal relations between $\leff$ and $R$ also implicitly assume this flat-in-$\leff$ prior.  Because $\leff$ is a strong function of the neutron star radius (\citealt{De2018,Raithel2018}), a flat distribution of $\leff$ implies a distribution of radii that strongly favors large radii. Similarly, a flat prior distribution in the spectral indices of an EOS predicts a distribution of radii and tidal deformability that increases with larger values (see, e.g., Fig. 2 of \citealt{Jiang2020}). Because $L_0$ also depends on $R_{1.4}$ \citep{Lattimer2001}, a flat distribution in $L_0$ implies a non-uniform distribution in $R$ and $\leff$ as well. Because of these non-linear relationships, flat priors in $R$, $\leff$, $L_0$, and the spectral EOS indices are fundamentally incompatible with one another.

\begin{figure}[ht]
\centering
\includegraphics[width=0.45\textwidth]{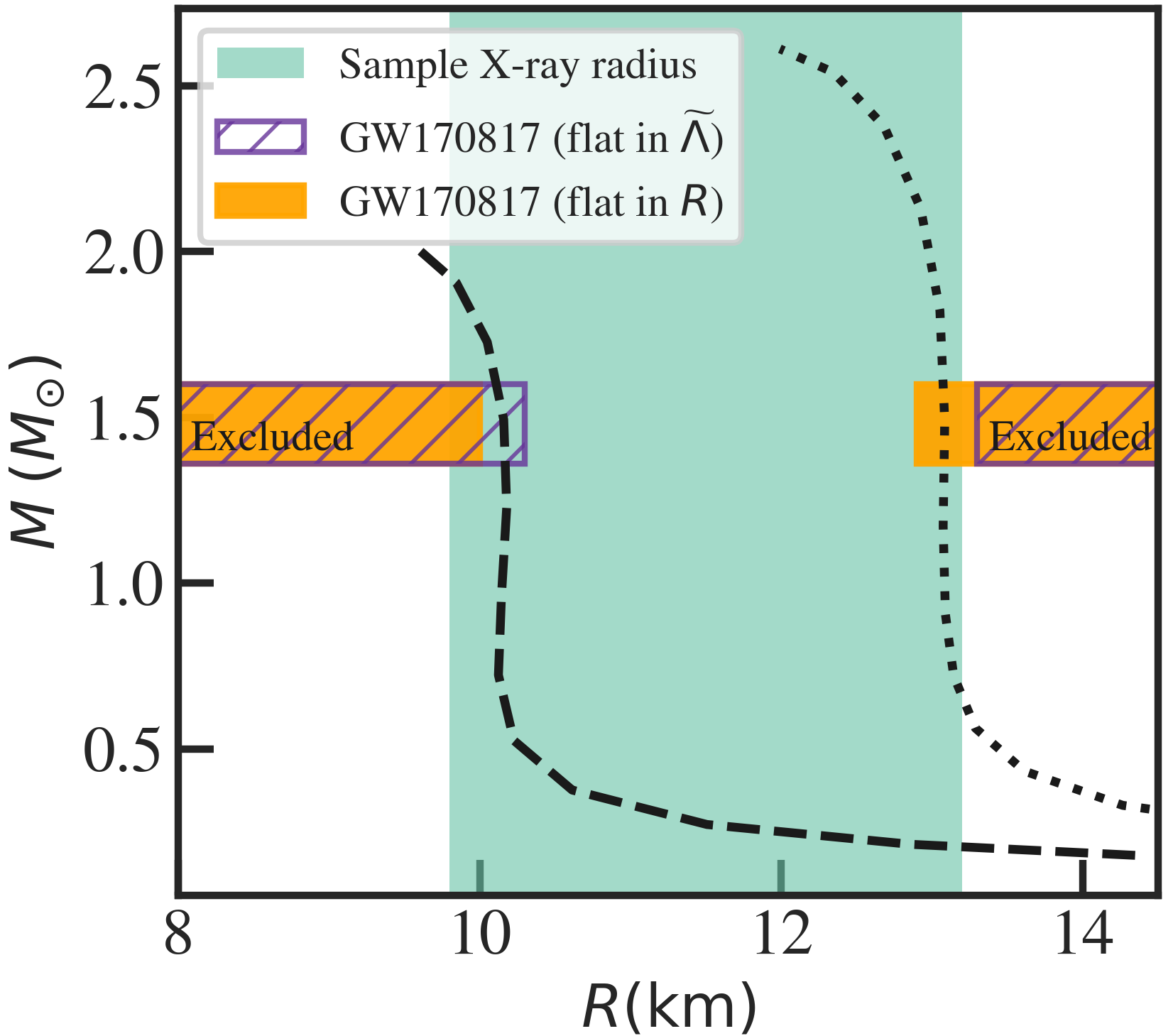}
\caption{\label{fig:EOSobs} Example consistency check between two theoretical EOS (shown as dashed and dotted lines) and a collection of astrophysical data. A fictitious, broad radius constraint from an X-ray observation is shown in green. We also show the 90\%-confidence exclusion bounds on the radius that are inferred from the LIGO posteriors on $\leff$   \citep{Abbott2019}, with two different choices of priors. In the purple hatched band, we show the radii that are excluded if one assumes a flat prior in $\leff$, as was used by the LIGO-Virgo collaboration. The orange band shows the radii that are excluded if a flat prior in radius is instead assumed (we will derive these bounds in $\S$\ref{sec:GWtransform}).  If the X-ray radius measurement, which assumes a flat-in-$R$ prior, and the raw gravitational wave constraint, which assumes a flat-in-$\leff$ prior, are combined, then the dashed EOS would be ruled out at 90\%-confidence, while the dotted EOS would be allowed. In contrast, if we require consistent priors that are both defined as flat-in-$R$ (green and orange), then the dotted EOS would be rejected, while the dashed EOS would be consistent with both measurements.}
\end{figure}

Combining posteriors that assume inconsistent priors can lead to biased conclusions, which we demonstrate with a simple example in Fig.~\ref{fig:EOSobs}. We start by requiring consistency between a particular theoretical EOS (shown as either dashed or dotted lines, for two sample EOS) and a set of two archival posteriors on the radius. In this example, one set of posteriors comes from a fictitious X-ray radius measurement (shown in green; assumes a flat-in-$R$ prior) and the second set of posteriors comes from GW170817 (excluded 90\% bounds shown with the purple hatched band; derived assuming a flat-in-$\widetilde{\Lambda}$ prior). When combining these inconsistent priors, we would falsely conclude that the dashed EOS is incompatible with the joint data at the 90\% confidence level, while the dotted EOS is allowed. However, if we instead transform the GW170817 radius constraint to assume a prior that is consistent with the X-ray measurement (i.e., flat-in-$R$; shown in orange), we find that, in fact, the dotted EOS is ruled out by the joint posteriors, while the dashed EOS is allowed, at 90\% confidence. 

In this example, the 90\% threshold for determining consistency between a dataset and the EOS is somewhat arbitrarily chosen (though, this criterion is often used in the EOS literature, as in, e.g., \citealt{Christian2020}). In order to be more quantitative in our comparison, we can also calculate the Bayesian evidence for each case of priors. For the case of the inconsistently-combined priors (flat-in-$\leff$ priors for GW170817 and flat-in-$R$ for the X-ray measurement), the Bayesian evidence favors the dotted EOS with a Bayes factor of 2.0. However, for the consistent set of flat-in-$R$ priors, it is the dashed EOS which is favored, with a Bayes factor of 2.1. While a swing of the odds ratio by a factor of $\sim$4 is of marginal significance, this cartoon example illustrates that adopting inconsistent priors can lead to opposing conclusions. Similar false conclusions can be drawn whether the consistent set of priors is ultimately defined in the radius domain, the $\widetilde{\Lambda}$-domain, or some other domain altogether.  Clearly, defining priors self-consistently is necessary for deriving unbiased constraints.

While this conclusion may be a general feature of Bayesian statistics, our goal in this paper is to demonstrate the importance of maintaining self-consistent priors and provide a method for doing so, specifically within the context of the neutron star literature, where it remains common to combine inconsistent priors and where the definition of a ``non-informative" prior varies across subfields. Most work highlighting the role of priors in constraining the neutron star EOS has focused on Bayesian inference schemes, and hence focuses on priors that are defined within the EOS domain \citep[e.g.,][]{Steiner2016,Carney2018}. Here, we focus on priors that are defined in external, observable domains, specifically for use in transforming the archival posteriors on $R$, $L_0$ and $\leff$ that are widely used for cross-domain comparisons.

\section{Defining Bayesian priors}
\label{sec:Bayes}

We start with a general review of Bayesian statistics, in order to illustrate the problems that can arise when performing cross-domain comparisons of archival posteriors. Bayes' theorem states that, when modeling some collection of data with a set of parameters $\vec{\theta}$, the posterior distribution on $\vec{\theta}$ is given by
\begin{equation}
\label{eq:Bayes}
P(\vec{\theta} | \mathrm{data}) \propto P_{\rm{pr}}(\vec{\theta} ) \mathcal{L}(\mathrm{data} |\vec{\theta} ),
\end{equation}
where $P_{\rm{pr}}(\vec{\theta})$ represents the Bayesian prior on  $\vec{\theta}$ and
$\mathcal{L}(\mathrm{data} | \vec{\theta})$ represents the likelihood of observing the measured data given a particular set of values for $\vec{\theta}$.

We can transform this measurement of $\vec{\theta}$ to a new set of parameters, $\vec{\phi}$, with a simple transformation of variables,
\begin{equation}
\label{eq:transform}
P(\vec{\phi} | \mathrm{data})  = P(\vec{\theta} | \mathrm{data}) \mathcal{J}\left( \frac{\vec{\theta}}{\vec{\phi}} \right),
\end{equation}
where $\mathcal{J}$ represents the Jacobian of transformations. In the case that $\vec\theta$ and $\vec\phi$ are both single parameters, the Jacobian is simply $\rvert\partial \theta/\partial \phi \rvert$. From eq.~(\ref{eq:transform}), it is clear that, depending on the nature of this Jacobian, even a broad posterior on $\vec{\theta}$ can potentially lead to stringent constraints on $\vec{\phi}$, simply by the transformation of variables. While this is a general feature of Bayesian priors \citep[e.g., Chapter 1.3 of][]{Box1992}, it becomes particularly relevant for neutron star EOS inferences, in which individual measurements are often quite broad, and in which key parameters of interest -- namely, $L_0$,  $R$, and $\leff$ -- are high-powered functions of one another. As a result, requiring simultaneous consistency between a theoretical EOS and a set of posteriors that assume flat priors in more than one of these domains (as in e.g., \citealt{Krastev2019,Lim2019,Blaschke2020,Christian2020,Fattoyev2020,Khanmohamadi2020,Marczenko2020}), can potentially lead to false conclusions (as shown in Fig.~\ref{fig:EOSobs}).

In order to avoid such biases, it is necessary to decide, a priori, what one wants to take as ``known" about the population of neutron stars or about the behavior of nuclear matter, and then define priors in the other domains accordingly. There is freedom to choose the domain in which the initial set of priors is defined, but once that choice is made, it fixes the priors for the other variables. In the following, we provide transformation functions that can be used to map a prior on $L_0$, $R$, or $\leff$ to the other domains, for use in self-consistent cross-domain comparisons.

We start with priors that are defined with respect to the slope of the nuclear symmetry energy. If we consider $L_0$ to be the fundamental variable on which we want to define the prior, then we can define corresponding priors on $R$ and $\leff$ according to
\begin{subequations}
\label{eq:priors_P}
\begin{equation}
 P_{\rm pr;~L_0}(L_0) = P_{\rm pr}(L_0)
\end{equation}
\begin{equation}
 P_{\rm pr;~L_0}(R) = P_{\rm pr}(L_0) \biggr \rvert \frac{\partial R}{\partial L_0}  \biggr \rvert^{-1}
\end{equation}
\begin{equation}
\label{eq:prLambda_P}
 P_{\rm pr;~L_0}(\leff) = P_{\rm pr}(L_0)  \biggr \rvert \frac{\partial R}{\partial L_0}  \biggr \rvert^{-1} \biggr\rvert \frac{\partial \leff}{\partial R} \biggr \rvert ^{-1}.
\end{equation}
\end{subequations}
In these equations, we have introduced a short-hand notation for the prior, $P_{\rm pr;~X}(Y)$, which indicates a Bayesian prior on the measurement of a variable $Y$ that is defined with respect to a given prior on $X$. In defining the transformation of variables, we have chosen to expand the derivatives so that we ultimately have only two derivatives to calculate:  $\partial R/\partial L_0$ and $\partial \leff/\partial R$. This choice is particularly convenient because functions for $R(L_0)$ and $\leff(R)$ have been previously reported in other works, as we will review in $\S$\ref{sec:transform}.

If we instead choose the radius as the fundamental variable over which to define the prior, then the corresponding priors on the gravitational wave and nuclear parameters are given by
\begin{subequations}
\label{eq:priors_R}
\begin{equation}
 P_{\rm pr;~R}(L_0) = P_{\rm pr}(R) \biggr \rvert \frac{\partial R}{\partial L_0} \biggr \rvert 
\end{equation}
\begin{equation}
 P_{\rm pr;~R}(R) = P_{\rm pr}(R) 
\end{equation}
\begin{equation}
\label{eq:prLambda_R}
 P_{\rm pr;~R}(\leff) = P_{\rm pr}(R)  \biggr\rvert \frac{\partial \leff}{\partial R} \biggr \rvert^{-1},
\end{equation}
\end{subequations}
where a natural choice for a minimally-informative prior might be a bounded uniform distribution on $R$.

For the sake of completeness, we also include the set of self-consistent priors that are defined with respect to $\leff$,
\begin{subequations}
\label{eq:priors_Lambda}
\begin{equation}
 P_{\rm pr;~\leff}(L_0) =  P_{\rm pr}(\leff)  \biggr\rvert \frac{\partial \leff}{\partial R} \biggr \rvert \biggr \rvert \frac{\partial R}{\partial L_0} \biggr \rvert 
\end{equation}
\begin{equation}
 P_{\rm pr;~\leff}(R) = P_{\rm pr}(\leff)  \biggr\rvert \frac{\partial \leff}{\partial R} \biggr \rvert
\end{equation}
\begin{equation}
 P_{\rm pr;~\leff}(\leff) = P_{\rm pr}(\leff).
\end{equation}
\end{subequations}
For the flat prior on $\leff$ that the LIGO-Virgo collaboration assumed for GW170817 \citep{Abbott2017a,Abbott2019}, eq.~(\ref{eq:priors_Lambda}) represents the corresponding set of priors that are implied for $R$ and $L_0$.

Finally, we note that, in this paper, we focus on prior distributions that are flat in the variable of interest, in order to be consistent with the many published measurements that employ a flat prior in either $R$, $\leff$, or $L_0$. Lacking other information on what the distributions for these parameters should be, a flat distribution is a reasonable choice. A truly uninformative prior -- e.g., the Jeffreys' prior, for which the posterior is invariant to transformations of the prior \citep{Jeffreys1946} -- might be a more robust choice. However, the Jeffreys' prior is only well-defined for a particular experiment. For example, a Jeffreys' prior on $\leff$ can be derived from the Fisher information matrix for a gravitational wave measurement. The posterior for such a measurement will be invariant under a transformation of the Jeffreys' prior to, say, the radius, which can also be used to parameterize the strain data. But, the Jeffreys' prior for $R$ that is derived from the gravitational wave data will not be the same as the Jeffreys' prior for $R$ that would be derived from an X-ray measurement, which involves a different Fisher information matrix. In other words, there is no ``global" Jeffreys' prior that can be defined for independent experiments that measure the radius, tidal deformability, and symmetry energy. Thus, even if one were to adopt the Jeffreys' prior for one domain of interest, eqs.~(\ref{eq:priors_P})-(\ref{eq:priors_Lambda}) would still be necessary to transform that prior to the other domains.

\section{Transformation functions}
\label{sec:transform}
We now turn to deriving the transformation functions needed to calculate the priors in eqs.~(\ref{eq:priors_P})-(\ref{eq:priors_Lambda}).  For each transformation function, we make use of the appropriate relationships and provide the corresponding derivatives.

\begin{figure*}[ht]
\centering
\includegraphics[width=0.95\textwidth]{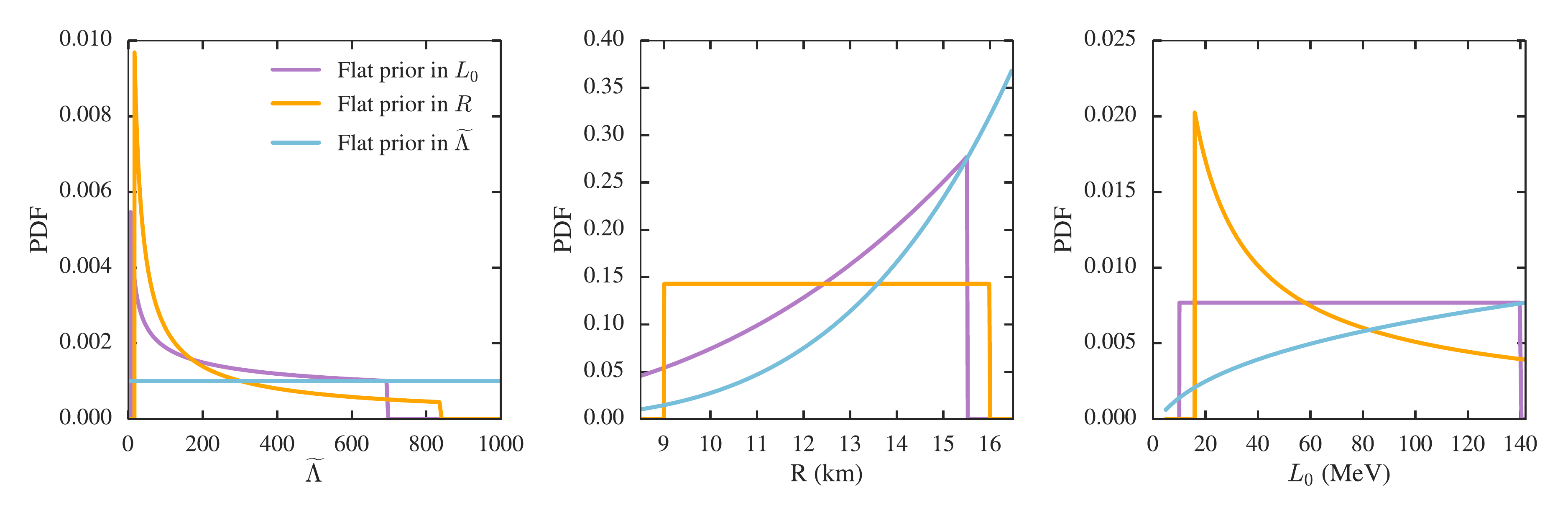}
\caption{\label{fig:priors} \textit{Left}: Prior distributions mapped to the domain of $\leff$. \textit{Center}: Prior distributions mapped to the domain of $R$. \textit{Right}: Prior distributions mapped to the domain of $L_0$. The purple lines represent the case of a uniform prior in $L_0$, which has been transformed to each of the domains using eq.~(\ref{eq:priors_P}). The orange lines represent a uniform prior in $R$, which has been transformed according to eq.~(\ref{eq:priors_R}). The blue lines represent a uniform prior in $\leff$, after transformation according to eq.~(\ref{eq:priors_Lambda}).}
\end{figure*}

\subsection{From the nuclear symmetry energy to the neutron star radius}

We start at the microscopic level, with the mapping between the the slope of the nuclear symmetry energy, $L_0$, and the neutron star radius. Many previous studies have found evidence of strong correlations between these parameters \citep[e.g.,][]{Lattimer2001,Steiner2013,Alam2016}. Here, we use the approximate relation
\begin{equation}
R_{1.4} \simeq (4.51 \pm 0.26) \left(\frac{L_0}{\rm MeV}\right)^{1/4} \mathrm{km},
\end{equation}
which was calculated as a function of pressure for a sample of realistic EOS in \citet{Lattimer2013} and later translated to be a function of $L_0$ in \citet{Tews2017}.
The derivative is then simply
\begin{equation}
\frac{\partial R}{\partial L_0} \simeq (1.128 \pm 0.065) \left(\frac{ L_0}{\rm MeV} \right)^{-3/4} ~\frac{\mathrm{km}}{\mathrm{MeV}},
\end{equation}
where we have assumed $R \approx R_{1.4}$, as is reasonable for EOS with nearly vertical mass-radius relations.

We note that, while the correlation between $L_0$ and the radius is strongest for smaller mass stars ($M \sim 1~\Ms$), a significant correlation between $L_0$ and $R_{1.4}$ is still present (see, e.g., Fig.~8 of \citealt{Fortin2016} for empirical $L_0-R$ correlations at different masses). Furthermore, because the neutron star mass distribution has a peak at $\sim$1.4~$\Ms$ \citep{Antoniadis2016}, the relationship between $L_0$ and $R_{1.4}$ is the more astrophysically-relevant correlation and, thus, is the correlation we focus on in this work. 

\subsection{From tidal deformability to the neutron star radius}

We now turn to the relationship between the radius and the effective tidal deformability measured from a gravitational wave event. Several studies have shown that $\leff$ is effectively a mono-parameteric function of the neutron star radius \citep{De2018,Raithel2018}, which scales quite strongly as $\leff \sim R^{5-6}$, where the exponent varies according to the slightly different assumptions made in these analyses. We use the formalism of \citet{Raithel2018} to exactly calculate $\partial \leff/\partial R$ below.

In that study, we used a quasi-Newtonian framework for calculating $\leff$, in which
\begin{equation}
\label{eq:leff}
\leff \approx \leff_{0} \left[ 1 + \delta_{0} (1-q)^2\right] + \mathcal{O}\left((1-q)^3\right),
\end{equation}
where
\begin{equation}
\label{eq:coef}
\leff_{0}  = \frac{15-\pi^2}{3 \pi^2} \xi^{-5} (1-2 \xi)^{5/2},
\end{equation}
\begin{equation}
\label{eq:correction}
\delta_{0} =  \frac{3}{104}(1-2 \xi )^{-2}\left(-10 + 94 \xi  - 83 \xi^2 \right),
\end{equation}
and $\xi$ was introduced as an effective compactness, defined as
\begin{equation}
\label{eq:xi}
\xi \equiv \frac{2^{1/5} G \mc}{ R c^2}.
\end{equation}
In these equations, $\mc$ is the chirp mass, $q$ is the mass ratio of the binary (defined such that $q\le1$), $G$ is the gravitational constant, and $c$ is the speed of light. Combining these results, one finds that the radius-dependence of the binary tidal deformability scales approximately as $\leff \sim R^6$. 

In this framework, the derivative of $\leff$ is then given by
\begin{equation}
\label{eq:fulldLamdR}
\frac{\partial \leff}{\partial R} \approx \frac{\partial \leff_0}{\partial R} 
	\left\{ 1 + \left[ \delta_0 + \leff_0 \left(\frac{\partial \delta_0}{\partial R} \right) \left(\frac{\partial \leff_0}{\partial R}\right)^{-1} \right] (1-q)^2 \right\},
\end{equation}
where we neglect the higher-order terms and we use the auxillary derivatives given by
\begin{equation}
\frac{\partial \delta_0}{\partial R} = -\frac{\delta_0 \xi}{R} \left[ \frac{ 54 + 22 \xi}{-10 + 114 \xi - 271 \xi^2 + 166 \xi^3} \right]
\end{equation}
and
\begin{equation}
\label{eq:dLeff0dR}
\frac{\partial \leff_0}{\partial R} = \frac{5 \leff_0 \xi}{R} \left( \frac{1}{\xi} + \frac{1}{1-2\xi} \right).
\end{equation}
The importance of the 2nd-order correction term in eq.~(\ref{eq:fulldLamdR}) increases with the chirp mass, $\mc$, and with the mass asymmetry of the binary. That is, larger values of $\mc$ and smaller values of $q$ will both act to increase the coefficient of the 2nd-order term. However, even for a very large $\mc=1.44~\Ms$, as was measured for GW190425, and for $q=0.7$, as was the lower limit for both GW170817 and GW190425, the correction term is at most 4\%. Thus, we neglect the 2nd-order correction term and simply approximate
\begin{equation}
\label{eq:dLeffdR}
\frac{\partial \leff}{\partial R}  \approx \frac{\partial \leff_0}{\partial R},
\end{equation} 
which scales approximately as $R^5$, with only a weak dependence on the individual component masses. 

\begin{figure*}[ht]
\centering
\includegraphics[width=0.95\textwidth]{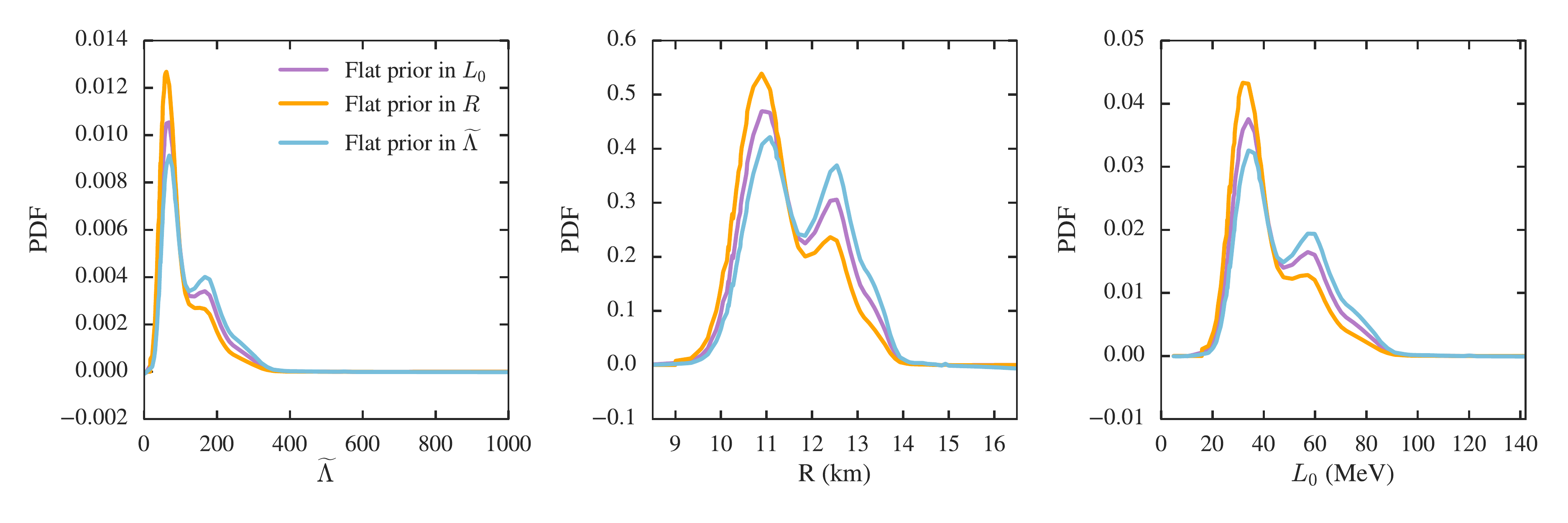}
\caption{\label{fig:GW170817} \textit{Left}: Posteriors on $\leff$ from GW170817, after reweighting for each set of priors and mapping to the chirp mass of event GW190425. The purple line indicates the case of a uniform distribution in $L_0$; the orange line represents a uniform prior in $R$; and the blue line represents a uniform prior in $\leff$, as was originally used by the LIGO-Virgo Collaboration. \textit{Center}: Constraints on $R$ inferred from each posterior on $\leff$. \textit{Right}: Constraints on $L_0$ inferred from each posterior on $\leff$. }
\end{figure*}
\begin{figure*}[ht]
\centering
\includegraphics[width=0.95\textwidth]{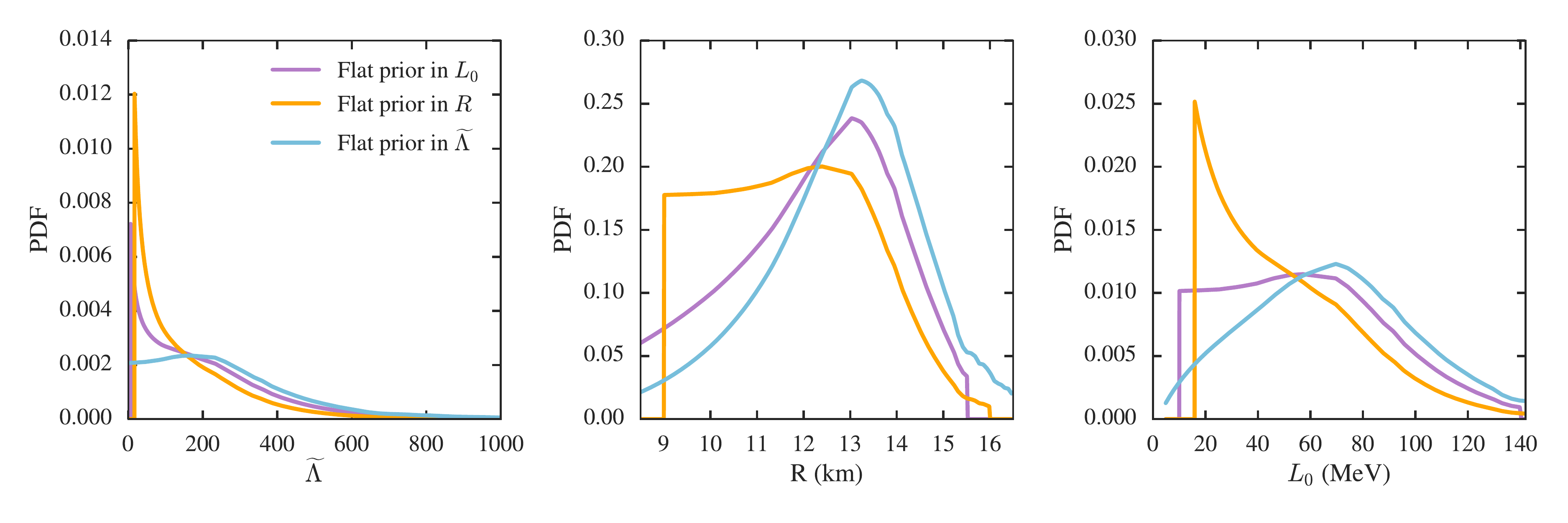}
\caption{\label{fig:GW190425} Same as Fig.~\ref{fig:GW170817}, but for the posteriors on $\leff$ measured from GW190425. The measurement of $\leff$ was much less significant for this event compared to GW170817, and thus the choice of prior strongly influences the subsequent inference of $R$ or $L_0$. In particular, for the original prior assumed by the LIGO-Virgo collaboration, the low-significance constraint on $\leff$ implies an artificial measurement of $R$.}
\end{figure*}

Finally, it is worth noting that, for the case of a neutron star-black hole binary (as is potentially consistent with event GW190814, reported in \citealt{Abbott2020}), the tidal deformability of the black hole is expected to be zero and, thus, the measured tidal parameter simply reduces to the tidal deformability of the neutron star. This simplification destroys the inherent symmetry that gave rise to the weak dependence on component masses in eq.~(\ref{eq:leff}). As a result, the tidal deformability from a neutron star-black hole binary will primarily probe the neutron star compactness, rather than the intrinsic radius (for further discussion, see \citealt{Raithel2018}).

\subsection{Summary of transformations}
We now apply these transformation functions to compute the priors in eqs.~(\ref{eq:priors_P})-(\ref{eq:priors_Lambda}). For each fundamental variable, we assume a bounded uniform distribution. We bound the uniform prior on $\leff$ to be positive and less than 1200, which is well above the limits that were derived for either GW170817 (with an adjusted chirp mass of $\mc=1.44~\Ms$)\footnote{The binary tidal deformability is a mass dependent quantity. In order to compare the results from GW170817 and GW190425 directly, we need to re-weight $\leff$ from the two events to have the same chirp mass. Thus, we adjust the chirp mass of GW170817 to match the central value of the chirp mass for GW190415, $\mc=1.44~\Ms$, in order to facilitate this comparison. For the chirp-mass adjusted posteriors on $\leff$ from GW170817, see \citet{LIGO2020}.} or GW190425. We bound the uniform prior on $R$ to be between 9 and 16~km, in order to broadly encompass all viable physics formulations and current measurements. Finally, we bound the uniform prior on the slope of the symmetry energy such that $L_0 \in [10,140]$~MeV, in order to be approximately consistent with a wide range of experimental results (for a review, see, e.g., \citealt{Lattimer2012}).

We show the resulting transformations of these priors in Fig.~\ref{fig:priors}. In blue, we show the original case of a uniform prior on $\leff$, as was used by the LIGO-Virgo collaboration for both GW170817 and GW190425. The middle panel shows how the flat prior in $\leff$ maps to a highly informative prior in $R$, which is biased towards large radii. The right panel shows that a flat prior in $\leff$ is moderately biased towards larger values of $L_0$. Figure~\ref{fig:priors} also shows how a uniform prior in $R$ or $L_0$ transforms to the other domains, in orange and purple lines, respectively. Clearly, a ``non-informative" prior in one domain can be highly informative in a different domain.  

Figure~\ref{fig:priors} also demonstrates the incompatibility of assuming flat priors in more than one of these domains. For example, a flat prior in $R$ assigns equal probability to stars with radii of 10 or 15~km, whereas a flat prior in $\leff$ assigns $8.5\times$ more probability to the larger star. A flat prior in $L_0$ implies that the 15~km star is $\sim3\times$ more likely than the 10~km star. While any of these may be a valid prior distribution to choose, they clearly describe very different physical assumptions.

\section{Example application to gravitational wave data}
\label{sec:GWtransform}

With these transformation functions in hand, we now turn to a concrete example, in order to further highlight how the interpretation of some specific measurements can rely on the priors. In this section, we will calculate posteriors for $\leff$ using priors that are minimally informative in either $\leff$, $R$, or $L_0$. We will then map each set of posteriors to constraints on $R$ and $L_0$, in order to illustrate the sensitivity of the resulting constraints to the particular choice of priors. 

We start with the posteriors on $\leff$ from GW170817, which were measured assuming a flat prior in $\leff$ \citep{Abbott2017a,Abbott2019}. These posteriors are shown in blue in the left panel of Fig.~(\ref{fig:GW170817}), for an adjusted chirp mass of $\mc=1.44~\Ms$. We then modify the published posterior to calculate the posterior that would have been inferred had the prior been uniform in radius (shown in orange) or uniform in $L_0$ (shown in purple). We calculate these new posteriors as
\begin{equation}
\label{eq:modifyPr}
P(\leff | \mathrm{data}) =P_{\rm old}(\leff | \mathrm{data}) \left[ \frac{P_{\rm pr,~new}(\leff)}{P_{\rm pr,~old}(\leff)}\right],
\end{equation}
where $P_{\rm pr,~new}(\leff)$ indicates the new prior, which is given by eq.~(\ref{eq:prLambda_P}) for the case of a uniform prior in $L_0$ or by eq.~(\ref{eq:prLambda_R}) for the case of a uniform prior in $R$ . Here, $P_{\rm pr,~old}(\leff)$ represents the original, uniform prior on $\leff$ and $P_{\rm old}(\leff | \mathrm{data})$ represents the original, published posterior. By dividing the reported posterior by the old prior, we recover the original likelihood.

We then transform each of the three, new posteriors on $\leff$ to find the corresponding constraints on $R$, according to 
\begin{equation}
P(R | \mathrm{data}) = P(\leff | \mathrm{data}) \biggr \rvert \frac{\partial \leff}{\partial R} \biggr \rvert.
\end{equation}
We similarly transform the posteriors on $\leff$ to constraints on $L_0$, according to
\begin{equation}
P(L_0 | \mathrm{data}) = P(\leff | \mathrm{data}) \biggr \rvert \frac{\partial \leff}{\partial R} \biggr \rvert \biggr \rvert \frac{\partial R}{\partial L_0} \biggr \rvert.
\end{equation}

The inferred constraints on $R$ and $L_0$ are shown in the middle and right panels of Fig.~\ref{fig:GW170817}, respectively. At 68\% confidence (highest-posterior density), the radius is constrained to  $R=10.9 \substack{+1.8\\-0.6}$~km for uniform priors in $L_0$, $R=10.9\substack{+0.8\\-0.7}$~km for uniform priors in $R$, and $R=11.1\substack{+1.8\\-0.6}$~km for uniform priors in $\leff$. There is a small difference between the inferred constraints, depending on which choice of prior is used. 
In particular, assuming a flat prior in $\leff$ or $L_0$ leads to evidence for slightly larger radii compared to the radii that are inferred when a flat prior distribution in $R$ is assumed. Accordingly, radius constraints that are derived from posteriors on $\leff$ which assume a flat-in-$\leff$ prior \citep[e.g.,][]{Annala2018,De2018,Raithel2018,Coughlin2019,Radice2019} will tend to favor larger radii, purely as an artifact of the prior. This effect will be important to take into account when comparing gravitational wave constraints on $R$ to X-ray constraints on $R$, which typically assume priors that are flat in the radius. However, the data for GW170817 are constraining enough that the overall effect of the prior remains small for this event. 

In contrast, Fig.~\ref{fig:GW190425} shows that the constraints inferred from $\leff$ for GW190425 are much more sensitive to the choice of the prior. As for GW170817, the LIGO-Virgo collaboration reported posteriors on $\leff$ assuming a uniform prior distribution on $\leff$ \citep{LIGO2020}. However, unlike GW170817, the resulting posteriors for GW190425 essentially represent a non-detection: the authors state that they lack the requisite sensitivity to detect matter effects for this system \citep{LIGO2020}. Nevertheless, they report constraints on $\leff$, the neutron star EOS, and $R$, assuming that GW190425 is indeed a binary neutron star system based on its component masses. Following suit, we re-weight the reported posteriors on $\leff$ to determine the posteriors that would have been inferred had a uniform prior in $R$ or $L_0$ instead been used, according to eq.~(\ref{eq:modifyPr}). The resulting posteriors, and their transformations to $R$ and $L_0$, are shown in Fig.~\ref{fig:GW190425}. 

We find that the choice of prior strongly influences the resulting constraints on $R$ and $L_0$ for GW190425. In particular, the assumption of a flat prior in $\leff$ leads to the inference of quite large radii, $R=13.2 \substack{+1.5\\-1.7}$~km (68\% credibility interval), even though no significant matter effects were detected in the actual measurement. The inference of large radii is purely an artifact of the transformation of variables. If we instead use a uniform prior in the radius, then the corresponding constraints on $R$ are also relatively uniform, as one would expect from a non-detection, such that it does not make sense to report a 68\% credibility interval. We find that the constraints on $R$ are essentially flat across the range of 9-13~km, with values of $R \gtrsim 13$~km disfavored.  Figure~\ref{fig:GW190425} thus demonstrates that the prior outweighs the actual data for this event.  Moreover, Fig.~\ref{fig:GW190425} demonstrates that comparing in the radius domain, when the measurement and original prior were defined in the $\leff$ domain, can produce inflated evidence for large radii, even in the absence of a measured signal. 

The conclusion that the prior outweighs the data for GW190425 may be obvious when the posteriors are examined in the domain in which they are made. In this case, the relatively flat posterior measured for $\leff$ is clearly mostly consistent with the flat prior that was assumed, and we can conclude that the event was not very informative. The picture becomes less clear, however, when transforming to a different domain and then making comparisons in that domain.  In fact, several studies are already starting to compare the radii inferred from GW190425, which are completely prior-dominated, to the predictions of theoretical EOS \citep[e.g.,][]{Blaschke2020,Marczenko2020}. Even when made only at a qualitative level, such comparisons provide false evidence for large radii, while the data themselves bear little-to-no constraining power.

The two gravitational wave events that have been detected so far are relatively straightforward to identify as ``strongly" and ``weakly" constraining events. However, in the coming years, it is likely that the LIGO-Virgo collaboration will measure many events whose constraints on $\leff$ fall in the more intermediate category of constraining power. When interpreting these events, it will be important to not only define priors that are consistent with one another, but to explicitly acknowledge which domain the priors are defined within and to understand how that choice influences subsequent transformations to other observable quantities.

\section{Composite constraints on the neutron star radius}
\label{sec:constraints}

With the new prescription for defining priors introduced in this paper, we now present summary constraints on the neutron star radius, using the latest results from X-ray data, gravitational waves, and nuclear constraints on $L_0$.

\begin{figure}[ht]
\centering
\includegraphics[width=0.45\textwidth]{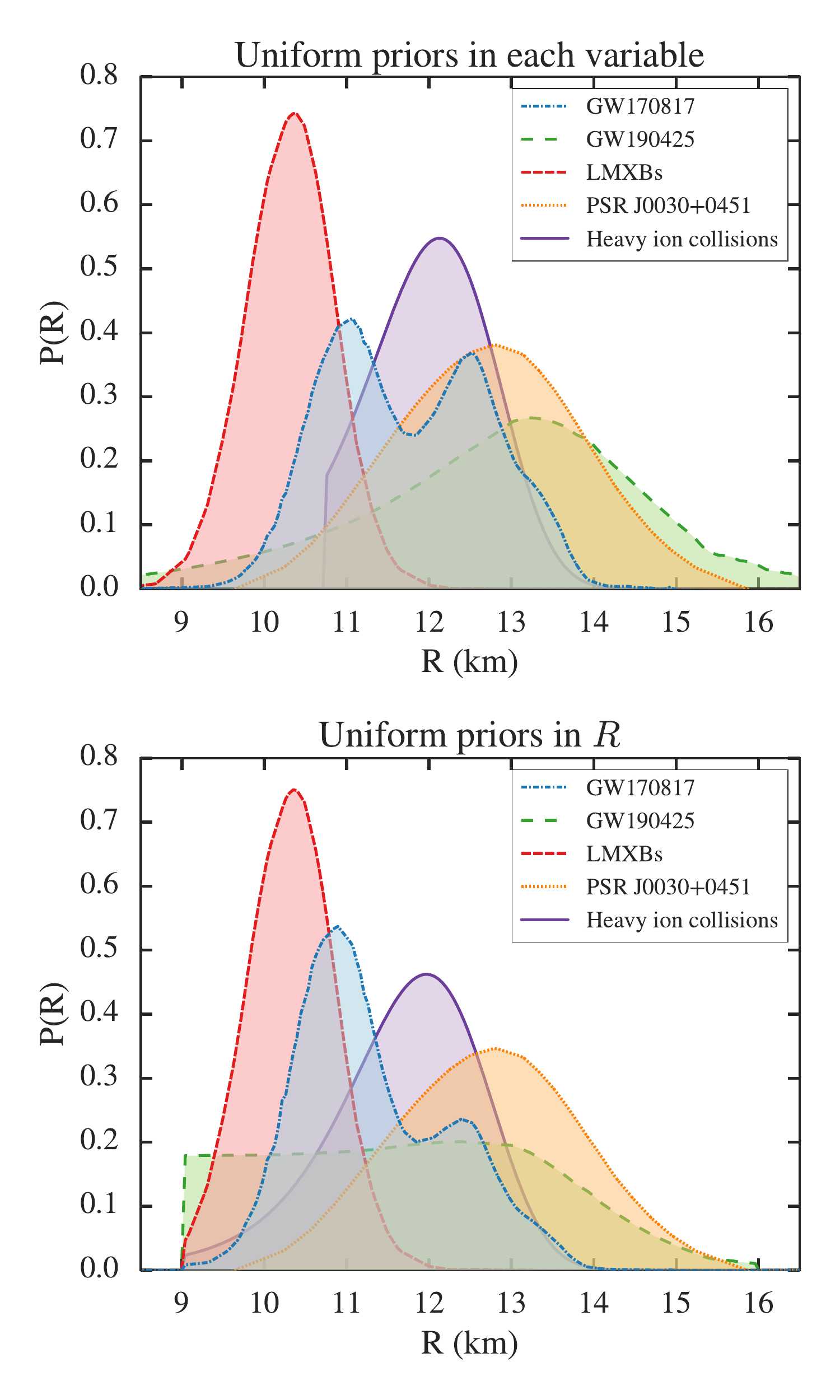}
\caption{\label{fig:pdfR} \textit{Top}: Constraints on the neutron star radius from X-ray observations, gravitational wave inference, and nuclear experimental data, assuming a uniform prior in each of the measured quantities (i.e., $\leff, R$, and $L_0$).  \textit{Bottom}: Constraints on the neutron star radius from the same data, but now assuming a uniform prior in the radius. We find that using prior distributions that are chosen to be minimally-informative in the radius results in more evidence for smaller radii.}
\end{figure}

These results are summarized in Figure~\ref{fig:pdfR} for two choices of priors. In each panel of this figure, we include likelihoods from GW170817 \citep[in blue,][]{Abbott2019} and GW190425 \citep[in green,][]{LIGO2020}. For X-ray radius constraints, we include the NICER analysis of PSR J0030+0451 \citep[in orange,][]{Riley2019}, as well as an analysis of 12 neutron stars in LMXBs \citep[in red, as analyzed by][]{Ozel2016a}. We could well have used a different set of LMXB radius constraints, such as those of \citet{Steiner2013}, which differ from the \citet{Ozel2016a} analysis in the physical assumptions used to interpret the observed fluxes from these sources. While the \citet{Steiner2013} analysis finds evidence of somewhat larger radii (10.6$-$12.6~km at 68\% confidence) than what is shown in Fig.~\ref{fig:pdfR}, the following conclusions would still hold for either set of LMXB data. Finally,  Figure~\ref{fig:pdfR} also includes a recent constraint on $L_0$ from an analysis of single and double ratios of neutron and proton spectra from heavy-ion collisions \citep[in purple,][]{Morfouace2019}. While we only include a single constraint on $L_0$, we note that this posterior ($L_0=49.6\pm13.7$~MeV, with values below 32~MeV or above 120~MeV forbidden\footnote{The constraints from \citet{Morfouace2019} assume a uniform prior for $32 < L_0 < 120$~MeV. Because the prior goes to zero outside of this range, we cannot rigorously recover the likelihood for very large or small values of $L_0$. Instead, when we re-weight the posterior to use a prior that is uniform in $R$, we simply assume that the likelihood continues as the inferred Gaussian outside of this range.}) is consistent with the results of a recent meta-analysis of several dozen studies that determined $L_0 = 58.7\pm28.1$~MeV \citep{Oertel2017}. Thus, we include the \cite{Morfouace2019} results in Fig.~\ref{fig:pdfR} as a representative and recent example of Bayesian constraints on $L_0$.

The top panel of Fig.~\ref{fig:pdfR} shows the composite posteriors on $R$ assuming uniform priors in the domain of each measurement, as is commonly done in the literature; i.e., uniform priors on $\leff$ for the gravitational wave events, uniform priors on $R$ for the X-ray data, and uniform priors on $L_0$ for the heavy-ion collision inference. In contrast, the bottom panel of Fig.~\ref{fig:pdfR} shows the constraints on $R$ that are derived when a uniform prior on $R$ is assumed for each measurement. Figure~\ref{fig:pdfR} illustrates that using a uniform prior in each variable leads to more evidence for larger radii from the $\leff$ and $L_0$ measurements, while the radius measurements that are made in directly in this domain remain relatively small. Thus, by mixing posteriors with inconsistent priors, the resulting constraints become muddled.  In contrast, when we define the priors self-consistently in the radius domain, the resulting constraints are overall shifted to slightly smaller radii and a clearer picture emerges. In particular, with the inclusion of a prior that is flat in $R$, the constraints from GW190425 now disfavor large radii, the constraints from GW170817 are strongly peaked at $\sim$11~km, and the constraints from the symmetry energy measurement also shift to smaller radii, albeit by a smaller degree. Thus, with the inclusion of a self-consistent set of priors defined in the radius-domain, the data overall seem to prefer smaller radii, compared to the top panel of Fig.~\ref{fig:pdfR}. This conclusion does not rely on a comparison to any one particular measurement, but is a result of the mapping between $\leff$, $L_0$, and $R$. One could also define the priors self-consistently with respect to $\leff$ or $L_0$ or even a different parameter altogether, in which case the inferred radii may shift to slightly larger values, compared to the bottom panel of Fig.~\ref{fig:pdfR}, but would, again, provide an internally consistent picture.

\begin{figure}[ht]
\centering
\includegraphics[width=0.45\textwidth]{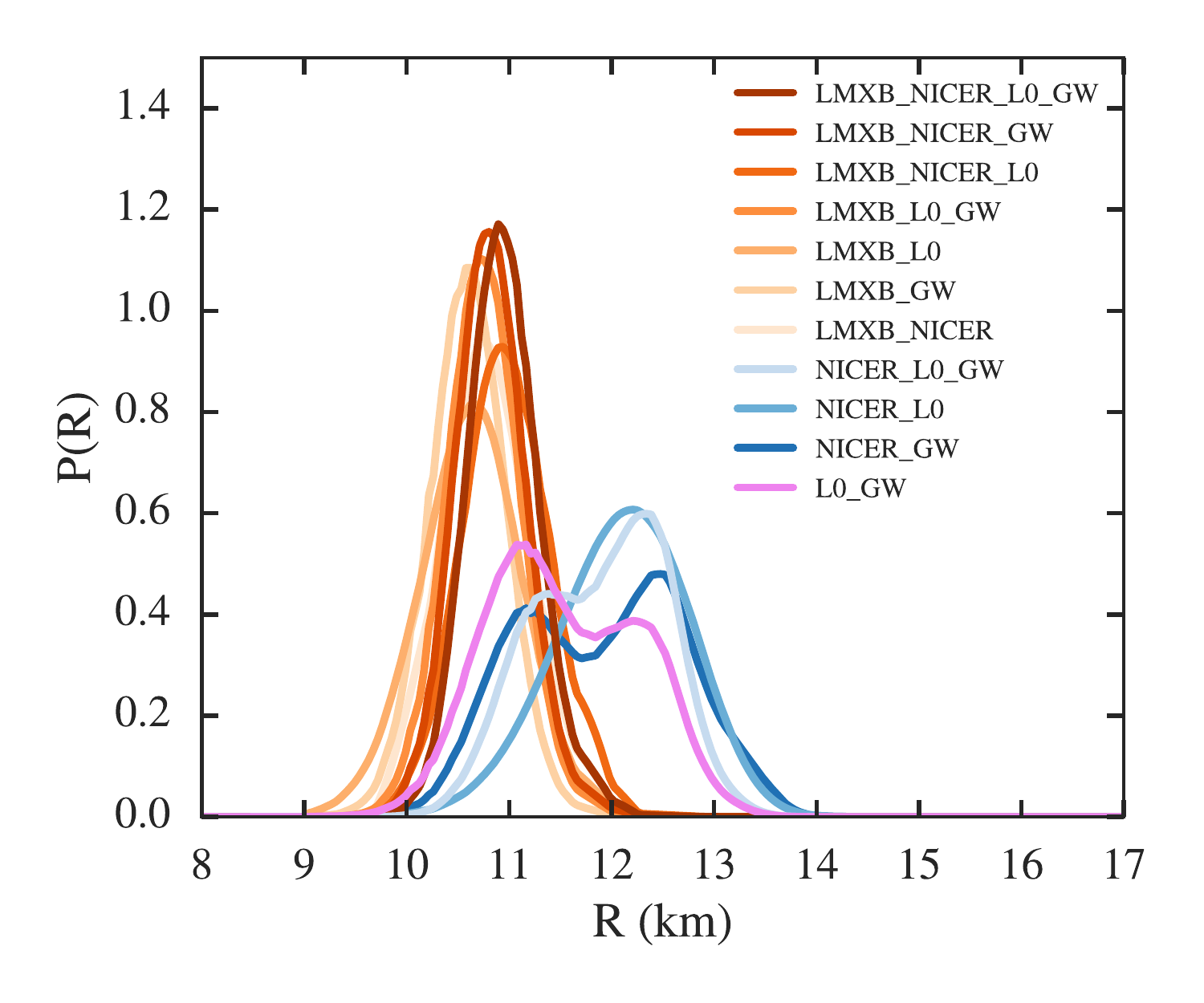}
\caption{\label{fig:jointPDF} Joint posterior distribution on the radius, determined by various combinations of experimental data. Orange lines correspond to any combination of experimental results that include the 12 LMXB sources. Blue lines indicate combinations that include the NICER source, PSR J0030+0451, as the only X-ray data.  The purple line shows the constraints inferred from only gravitational wave and nuclear contraints; i.e., with no X-ray data. }
\end{figure}

Finally, Fig.~\ref{fig:jointPDF} shows the joint posterior distribution for various combinations of these experimental and observational constraints, with uniform priors defined in the radius domain. The orange lines show the joint posterior distributions for any combination of experimental constraints that include the data for the 12 LMXB sources, with the darkest orange line representing the joint posterior including all of the data shown in Fig.~\ref{fig:pdfR}. The blue lines represent the joint posteriors for any combination of data that include the NICER pulsar as the only X-ray source. Finally, the purple line represents the joint posterior for just the nuclear and the gravitational wave constraints (i.e., excluding all X-ray sources).  

In these joint posteriors, we include just one analysis for each source or set of sources. If one wanted to include multiple analyses of the same data -- e.g., the two analyses of PSR J0030+0451 by \citet{Riley2019} and \citet{Miller2019} -- these could be incorporated as a weighted average of the posteriors, where the weight is an effective prior representing the confidence one has in each particular analysis. In this way, the total probability for the measurement still integrates to unity,  across all different analyses.

We find that the data from the 12 LMXB sources (as analyzed by \citealt{Ozel2016a}) are the most constraining measurement included in this paper. This is, in part, a result of the large number of neutron stars included in that analysis. While the strength of the LMXB data varies by source (see Fig.~11 of \citealt{Ozel2016a}), in every case, the data are more constraining than the very broad, flat-in-$R$ prior that was used. We find that any joint posteriors that contain these narrowly-peaked LMXB data point to $R\sim10-11.5$~km. Moreover, small radii are supported by \textit{any} combination of the results that exclude the NICER data, including the combination of gravitational wave and $L_0$ constraints alone.  In contrast, if the NICER source is included as the only X-ray data, then the resulting radii are much larger, $R\sim12-13$~km. Currently, the NICER collaboration has published radius constraints for just a single source, PSR J0030+0451, using a multi-component, phenomenological pulse-profile model to fit the data.  As more physical pulse-profile models are developed and more sources are included in the analysis, it will be interesting to see whether this systematic offset persists. Finally, we note that, if one were to use the LMXB analysis of \citet{Steiner2013} instead of the \citet{Ozel2016a} results, this tension with the NICER results would be reduced, but not completely erased.

Given the current tension between the NICER constraints and these LMXB and gravitational wave constraints, it is all the more crucial to consider the role of the prior when combining these posteriors. Different choices of the prior -- either on exterior parameters like $L_0$, $R$, or $\leff$, as considered in this paper, or on the parameters of the EOS itself -- will provide a different relative weight to each of these measurements. Thus, by naively picking a particular prior, one may also be granting more constraining power to a particular type of experiment. Of course, if the chosen prior is well-motivated, then this is exactly what should happen. However, we raise the issue here to point out that -- for the current state of sparse, and sometimes conflicting, neutron star data -- the choice of even ``non-informative" priors can significantly affect the resulting analysis and should be not be adopted naively.

As the community continues to work towards ever-more stringent constraints on the neutron star radius, these joint posteriors can be helpful for understanding the relative constraining power of each additional measurement and how this compares to the information provided by the prior. Joint posteriors can also help to identify systematic offsets between different types of measurements. For a recent example of a comparison of joint posteriors in a fully Bayesian inference that also includes a treatment for systematic offsets, see \citet{Al-Mamun2020}. Finally, we note again that regardless of which data are included in any meta-analysis, defining the priors to be self-consistent is an important step towards getting unbiased constraints.

\section{Conclusions}
With the recent flood of multi-messenger constraints on the neutron star EOS, it is important to start identifying the statistical biases that enter into comparisons of these diverse data sets. In this paper, we have highlighted the importance of defining a consistent set of priors and of understanding the role that those priors play, when transforming to different domains. We introduced a general prescription for calculating consistent priors and derived the relevant transformation functions so that archival posteriors from different experiments can be robustly compared.

Using the example of GW170817 and GW190425, we showed that assuming a Bayesian prior that is ``non-informative" in $\leff$ leads to a highly-informative constraint on $R$, even in the absence of a measured signal. In particular, a flat prior in $\leff$ biases the resulting constraint on $R$ to large values, whereas with a flat prior in the radius provides evidence for slightly smaller radii.

As the community continues to collect more and higher quality data, the relative importance of the priors should diminish. We have already shown this for the case of radius  constraints inferred from $\leff$ for GW170817, for which the choice of prior does not strongly affect the resulting posterior. However, for gravitational waves in particular, we may see far more low-significance events than we do GW170817-like events. Thus, if we hope to use the future constraints on $\leff$ to compare with other radius measurements, it is important to account for the role of the assumed priors. 

As new events -- gravitational and otherwise -- continue to be observed, the general prescription introduced in this paper will facilitate increasingly stringent, and statistically robust, constraints on the neutron star EOS.

\acknowledgments This work was supported in part by Chandra Grant GO7-18037X. C.~R. gratefully acknowledges support from NSF Graduate Research Fellowship Program Grant DGE-1746060. 

\bibliography{gwthermal}
\bibliographystyle{apj}

\end{document}